\documentclass[12pt]{article}
\usepackage[dvips]{graphicx}
\usepackage{amsmath,amssymb}
\def\lromn#1{\uppercase\expandafter{\romannumeral#1}}

\def\Slash#1{{\ooalign{\hfil$#1$\hfil\crcr\hfil$/$\hfil}}}

\begin{document}
%%%%%%%%%%%%%%%%%%%%%%%%%%%%%%%%%%
%%%%%%%%%%% Title page %%%%%%%%%%%
%%%%%%%%%%%%%%%%%%%%%%%%%%%%%%%%%%
\begin{titlepage}
 \begin{center}

  \hfill UT-HET 051 \\
  \hfill IPMU-11-0031 \\
  \hfill \today

  \vspace{1cm}
  {\large\bf Testing Higgs portal dark matter \\
via $Z$ fusion at a linear collider} \\
  \vspace{1cm}

  Shinya Kanemura$^{(a)}$ \footnote{kanemu@sci.u-toyama.ac.jp},
  Shigeki Matsumoto$^{(b)}$ \footnote{shigeki.matsumoto@ipmu.jp}, \\
  Takehiro Nabeshima$^{(a)}$ \footnote{nabe@jodo.sci.u-toyama.ac.jp},
  and
  Hiroyuki Taniguchi$^{(a)}$ \footnote{taniguchi@jodo.sci.u-toyama.ac.jp} \\

  \vspace{1cm}

  $^{(a)}${\it Department of Physics, University of Toyama, Toyama 930-8555, Japan} \\
  $^{(b)}${\it IPMU, TODIAS, University of Tokyo, Kashiwa 277-8583, Japan} \\
  \vspace{1cm}

  \abstract{
We investigate the possibility of detecting dark matter at TeV scale linear colliders in the scenario where the dark matter is a massive particle weakly interacting only with the Higgs boson $h$ in the low energy effective theory (the Higgs portal dark matter scenario). The dark matter in this scenario would be difficult to be tested at the CERN Large Hadron Collider when the decay of the Higgs boson into a dark matter pair is kinematically forbidden. We study whether even in such a case the dark matter $D$ can be explored or not via the $Z$ boson fusion process at the International Linear Collider and also at a multi TeV lepton collider. It is found that for the collision energy $\sqrt{s}>1$~TeV with the integrated luminosity $1$~ab$^{-1}$, the signal ($e^{\pm}e^-\to e^{\pm}e^-h^\ast \to e^{\pm}e^-DD$) can be seen after appropriate kinematical cuts. In particular, when the dark matter is a fermion or a vector, which is supposed to be singlet under the standard gauge symmetries, the signal with the mass up to about 100~GeV can be tested for the Higgs boson mass to be 120~GeV.
  } 
 \end{center} 
\end{titlepage} 

%%%%%%%%%%%%%%%%%%%%%%%%%%%%%%%%%%%%
%%%%%%%%%%% Introduction %%%%%%%%%%%
%%%%%%%%%%%%%%%%%%%%%%%%%%%%%%%%%%%%
\section{Introduction}

Dark Matter is one of the biggest mysteries in present physics and astronomy. It has been established that more than one fifth of the energy density in our Universe is occupied by dark matter~\cite{Komatsu:2010fb}. If the essence of the dark matter is a kind of particle, it must be electrically neutral and must be weakly interacting. As it has turned out that neutrinos cannot be the candidate, the dark matter should necessarily be a new massive content in physics beyond the standard model (SM). A plausible candidate for the dark matter is therefore a weakly interacting massive particle (WIMP). According to the WMAP experiment~\cite{Komatsu:2010fb}, the mass of the WIMP dark matter is at the TeV scale or less. Various direct and indirect dark matter search experiments are currently being performed~\cite{CDMS-II}-\cite{FERMI-LAT} and planned~\cite{XMASS}-\cite{XENON100-sen}. Moreover, we may be able to directly produce the dark matter and to test it at collider experiments such as the CERN Large Hadron Collider (LHC) and future linear colliders.

The fact that the mass scale of the WIMP dark matter is similar to that of the electroweak symmetry breaking would indicate that there is a connection between the Higgs boson and the dark matter. There are many new physics models involving a dark matter candidate. In some of them, it can happen that the dark matter couples only to the Higgs boson in the low energy effective theory, where stability of the dark matter is guaranteed by an unbroken discrete symmetry~\cite{higgsportal-scalar1}-\cite{higgsportal-vector1}. Such a scenario is often called the Higgs portal dark matter scenario~\cite{Higgsportal}. 

In the scenario of the Higgs portal dark matter, a collider signal at the LHC is expected to come from the $W$ boson fusion process $pp \to jjW^\ast W^\ast \to jjh^\ast \to jjDD$, where $D$ represents the Higgs portal dark matter whose spin is either $0$, $1/2$, or $1$~\cite{Kanemura:2010sh}, while $j$ is a jet originating in an energetic quark. When the mass of $D$ is less than one half of that of $h$, the invisible decay process $h \to DD$ opens, so that the signal would be detectable after appropriate kinematic cuts~\cite{Eboli:2000ze} unless the coupling constant between $h$ and $D$ is too small. On the other hand, if the decay $h \to DD$ is not kinematically allowed, the detection of the signal would be hopeless for the dark matter which is consistent with the WMAP and direct detection experiments~\cite{Kanemura:2010sh}. 

In this Letter, we investigate the possibility whether the Higgs portal dark matter can be tested at TeV scale linear colliders such as the International Linear Collider (ILC)~\cite{Flacher:2008zq} and the Compact Linear Collider (CLIC)~\cite{Accomando:2004sz} even in the case where the decay $h \to DD$ is not kinematically allowed. In the case of $m_{D}^{}<m_{h}^{}/2$, the process $e^+e^- \to Zh^\ast \to ZDD$ has been studied for the collision with the center of mass energy of $\sqrt{s} = 350$~GeV~\cite{Matsumoto:2010bh}. We here study pair production processes of the dark matter via $Z$ boson fusion from electron-positron ($e^+e^-$) and electron-electron ($e^-e^-$) collisions. It is found that for the collision energy $\sqrt{s}>1$~TeV with the integrated luminosity $1$~ab$^{-1}$, the signal ($e^{\pm}e^-\to e^{\pm}e^-h^\ast \to e^{\pm}e^-DD$) could be seen even for $m_{D}^{}>m_{h}^{}/2$ after appropriate kinematical cuts, when the mass of $D$ is not much heavier than that of the $W$ boson, especially for the dark matter $D$ being a fermion or a vector.

%%%%%%%%%%%%%%%%%%%%%%%%%%%%%
%%%%%%%%%%% Model %%%%%%%%%%%
%%%%%%%%%%%%%%%%%%%%%%%%%%%%%
\section{The model}

We here consider the simple model in which a dark matter field is added to the SM. We impose an unbroken $Z_2$ parity, under which the dark matter is assigned to be odd while the SM particles are to be even. Stability of the dark matter is guaranteed by the $Z_2$ parity. We consider three possibilities for the spin of the dark matter; i.e., the real scalar $\phi$, the Majorana fermion $\chi$ and the real massive vector $V_{\mu}$. 

The Lagrangian for each case of the dark matter is given by 
\begin{eqnarray}
 {\cal L}_{\rm S}
 &=&
 {\cal L}_{\rm SM}
 +
 \frac{1}{2} \left(\partial \phi\right)^2
 -
 \frac{1}{2} M_{\rm S}^2 \phi^2
 -
 \frac{c_{\rm S}}{2}|H|^2 \phi^2
 -
 \frac{d_{\rm S}}{4!} \phi^4, 
 \label{eq:S} \\
 {\cal L}_{\rm F}
 &=&
 {\cal L}_{\rm SM}
 +
 \frac{1}{2}\bar\chi\left(i\Slash{\partial} - M_{\rm F}\right)\chi
 -
 \frac{c_{\rm F}}{2\Lambda} |H|^2 \bar\chi \chi
 -
 \frac{d_{\rm F}}{2\Lambda} \bar\chi \sigma^{\mu\nu} \chi B_{\mu\nu}, 
 \label{eq:F} \\
 {\cal L}_{\rm V}
 &=&
 {\cal L}_{\rm SM}
 -
 \frac{1}{4} V^{\mu\nu} V_{\mu \nu}
 +
 \frac{1}{2} M_{\rm V}^2 V_\mu V^\mu
 +
 \frac{c_{\rm V}}{2} |H|^2 V_\mu V^\mu
 -
 \frac{d_{\rm V}}{4!} (V_\mu V^\mu)^2, 
 \label{eq:V}
\end{eqnarray}
where $M_i$($i = $ S, F and V) are the bare  masses of $\phi$, $\chi$ and $V_{\mu}$, $c_i$ and $d_i$ are dimensionless coupling constants, $\Lambda$ is a dimensionfull parameter, and $V_{\mu\nu}$ and $B_{\mu\nu}$ are Abelian field strength tensors. The last term in Eq.~(\ref{eq:F}) is expected to be small because this is induced at the one loop level. Hence, we neglect this term in the following analysis. In this case, the dark matter in Eqs.~(\ref{eq:S})-(\ref{eq:V}) only couples to the SM particles via the Higgs doublet field $H$: i.e., it is so-called the Higgs portal dark matter. 

After the electroweak symmetry breaking, the Higgs field $H$ obtains the vacuum expectation value $\langle H \rangle = (0,v)^T/\sqrt{2}$ with the value of $v$ being 246 GeV, and the physical mass of each dark matter particle is therefore given by 
\begin{eqnarray}
 m^2_{\phi}
 &=&
 M_{\rm S}^2
 +
 \frac{c_{\rm S}v^2}{2}, 
 \\
 m_{\chi}^{}
 &=&
 M_{\rm F}
 +
 \frac{c_{\rm F}v^2}{2\Lambda}, 
 \\
 m^2_{V}
 &=&
 M_{\rm V}^2
 +
 \frac{c_{\rm V}v^2}{2}. 
\end{eqnarray}
In our analysis, physical masses $m_{i}$ and coupling constants $c_{i}$ are treated as free parameters. Theoretical constraints and experimental bounds from the WMAP data and the direct search results on these models are discussed in Ref.~\cite{Kanemura:2010sh}.

%%%%%%%%%%%%%%%%%%%%%%%%%%%%%%
%%%%%%%%%%% Signal %%%%%%%%%%%
%%%%%%%%%%%%%%%%%%%%%%%%%%%%%%
\section{Dark Matter signals at the $e^+ e^-$collider}

\begin{figure}[t]
 \begin{center}
  \includegraphics[scale=0.7]{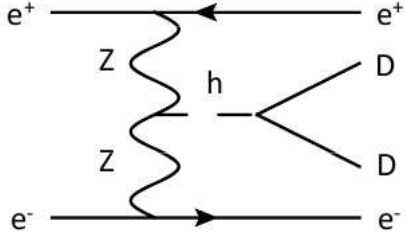}
  \caption{\small Feynman diagram of the signal process.}
  \label{fig:sig}
 \end{center}
\end{figure}

We consider the possibility to detect the dark matter at next generation of electron-positron linear colliders such as the ILC and the CLIC. In particular, we are interested in the case of $m_h < 2m_D$, where the Higgs boson cannot decay into a pair of dark matters. We concentrate on the $Z$ boson fusion process $e^+e^- \to e^+e^-Z^*Z^* \to e^+e^-h^* \to e^+e^-DD$ depicted in Fig.~\ref{fig:sig}. This process can, in principle, be used to detect the dark matter by measuring the outgoing electron and positron in the final state and by using the energy momentum conservation.

We impose the polarization for both incident electron and positron beams~\cite{Flacher:2008zq}; 
\begin{eqnarray}
\frac{N_{e^-_R}-N_{e^-_L}}{N_{e^-_R}+N_{e^-_L}} = 80\%,
\hspace{1cm}
\frac{N_{e^+_R}-N_{e^+_L}}{N_{e^+_R}+N_{e^+_L}} = 50\%,
\end{eqnarray}
where $N_{e^-_{R,L}}$ and $N_{e^+_{R,L}}$ are numbers of right (left) handed electron and positron in the beam flux per unit time. By using the polarized beams, the backgrounds which are mediated by the $W$ boson can be reduced. The backgrounds mediated by the $Z$ boson are reduced by the basic cut in Eq.~(\ref{eq:Minv}) as we will see soon.

\begin{figure}[t]
\begin{center}
\scalebox{0.7}{\includegraphics{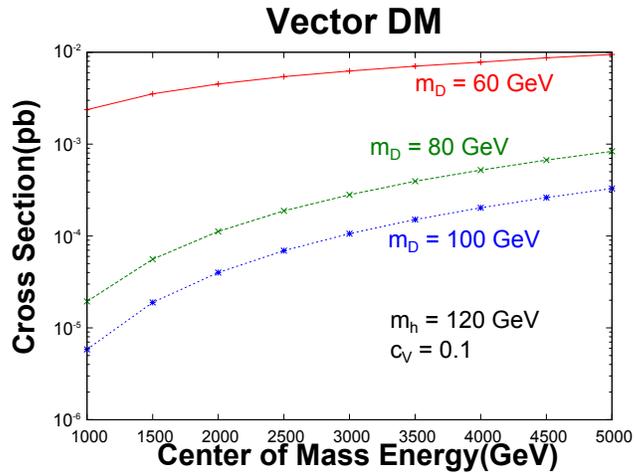}}
\caption{\small Cross section of the signal process $e^+e^-\to e^+e^-V_{\mu}V^{\mu}$ as a function of the center of mass energy $\sqrt{s}$. The mass of the dark matter $V_{\mu}$ is fixed to be 60, 80 and 100~GeV, while the coupling constant $c_{\rm V}^{}$ is taken to be 0.1.}
\label{fig:basic}
\end{center}
\end{figure}

The cross section of the signal process is the larger as the collision energy $\sqrt{s}$ increases, and its behavior is $\ln s$ as can be seen in Fig.~\ref{fig:basic}, so that the higher collision energy may be more huseful to detect the signal. However, for $\sqrt{s}= 1$-$5$~TeV, the outgoing electron and positron tend to be emitted to forward and backward directions, and the detectability of the leptons near the beam line is therefore essentially important. In this paper, we assume the detectable area as~\cite{Bambade:2006qc}
\begin{eqnarray}
 |\cos\theta|
 <
 0.9999416, 
 \label{eq:cos}
\end{eqnarray}
where $\theta$ is the scattering angle. Assuming the situation that the Higgs boson mass is already known, we impose the condition for the missing invariant mass $M_{\rm inv}$ as 
\begin{eqnarray}
 M_{\rm inv}
 >
 m_h^{}, 
 \label{eq:Minv}
\end{eqnarray}
in order to discuss the detection of the dark matter in the case $m_D^{} > m_h^{}/2$. The production cross sections of the signal process for $D=\phi$, $\chi$ and $V$ at the center of mass energy 1~TeV and 5~TeV are shown for $m_h = 120$~GeV in Table~\ref{table:event}. 

%%%%%%%%%%%%%%%%%%%%%%%%%%%%%%%%%%
%%%%%%%%%%% Parton Lv. %%%%%%%%%%%
%%%%%%%%%%%%%%%%%%%%%%%%%%%%%%%%%%
\section{Parton level signal and background}

Backgrounds against the signal process are all the process with the final state of $e^+e^-$ with a missing momentum. The main background processes are those with the final state $e^+e^-\nu_e\overline{\nu}_e$, $e^+e^-\nu_{\mu}\overline{\nu}_{\mu}$ and $e^+e^-\nu_{\tau}\overline{\nu}_\tau$. After the basic cuts given in Eqs.~(\ref{eq:cos}) and (\ref{eq:Minv}), the cross sections for $e^+e^-\to e^+e^-\nu_{e}\overline{\nu}_{e}$ and $e^+e^-\to e^+e^-\nu_{i}\overline{\nu}_{i}$ ($i = \mu$ or $\tau$) are evaluated as $1.15\times 10^{-1}$~pb and $8.87\times 10^{-4}$~pb at $\sqrt{s} = 1$~TeV, while they are $1.48\times 10^{-1}$~pb and $3.74\times 10^{-4}$~pb at $\sqrt{s} = 5$~TeV, respectively: see Table~\ref{table:event}. The signal to background ratio amounts to $10^{-3}$-$10^{-1}$ for the coupling constants being taken as $c_{\rm S}^{}=c_{\rm V}^{}=1$ and $c_{\rm F}^{}/{\Lambda}=0.1$~GeV$^{-1}$. In order to gain the signal significance we impose kinematical cuts as follows. 

\begin{figure}[p]
\begin{center}
\scalebox{0.29}{\includegraphics{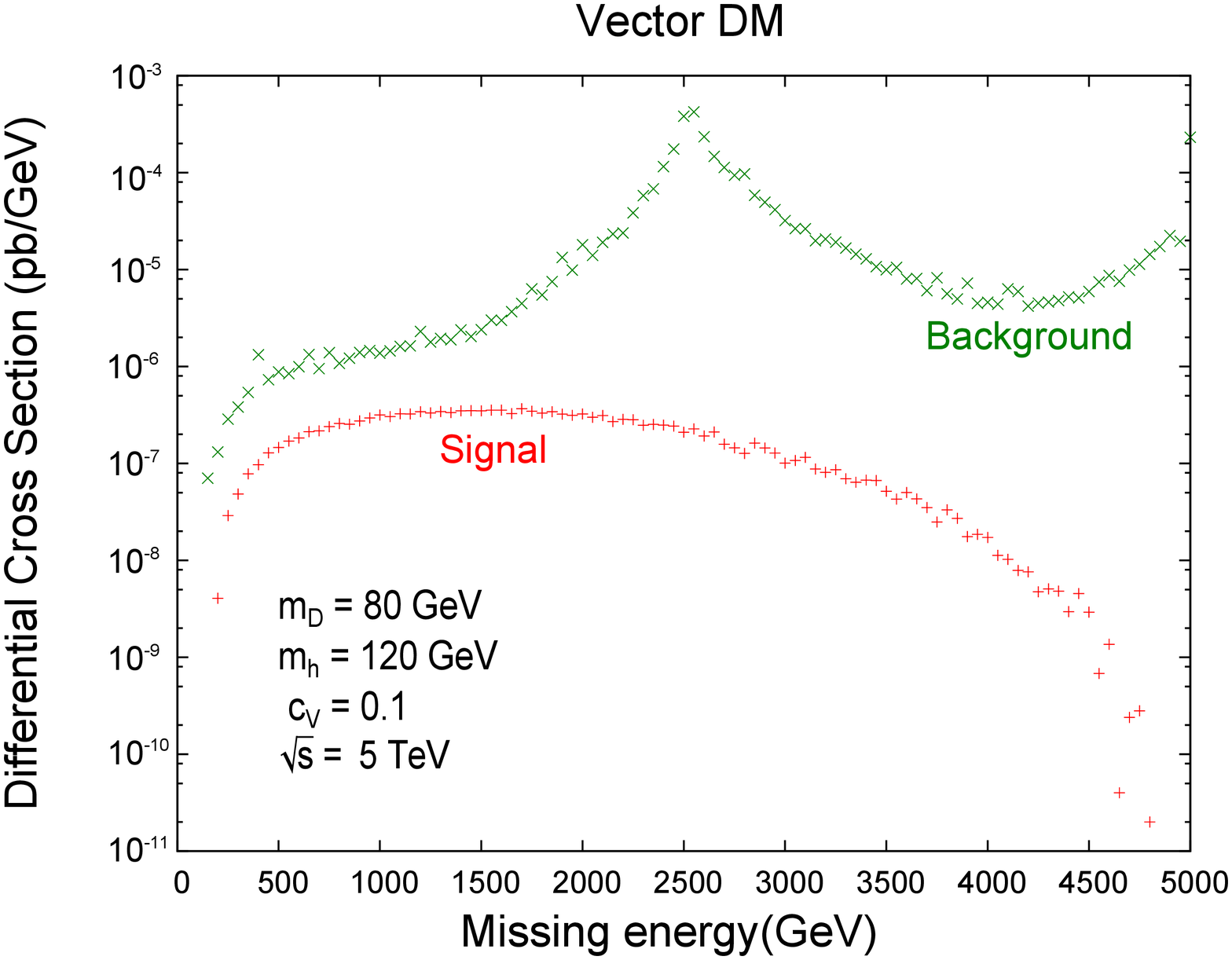}} \\
\scalebox{0.6}{\includegraphics{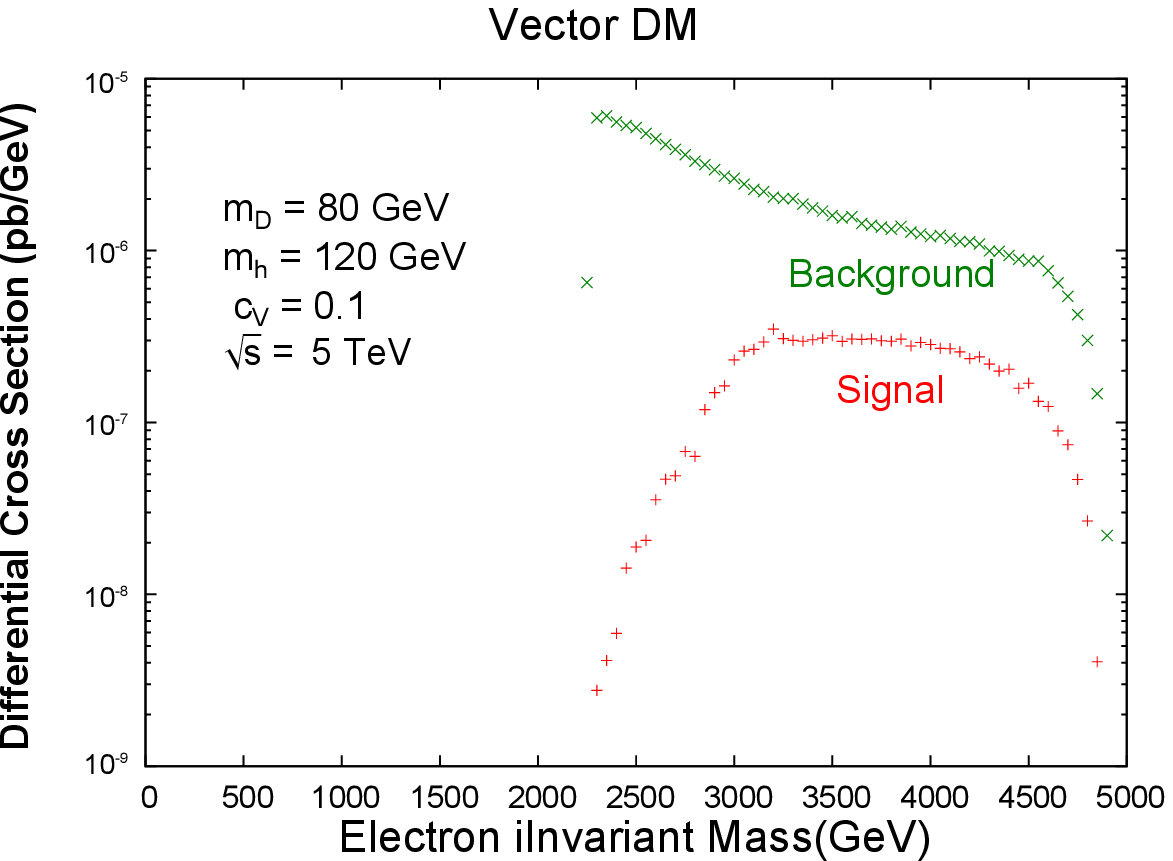}} \\
\scalebox{0.6}{\includegraphics{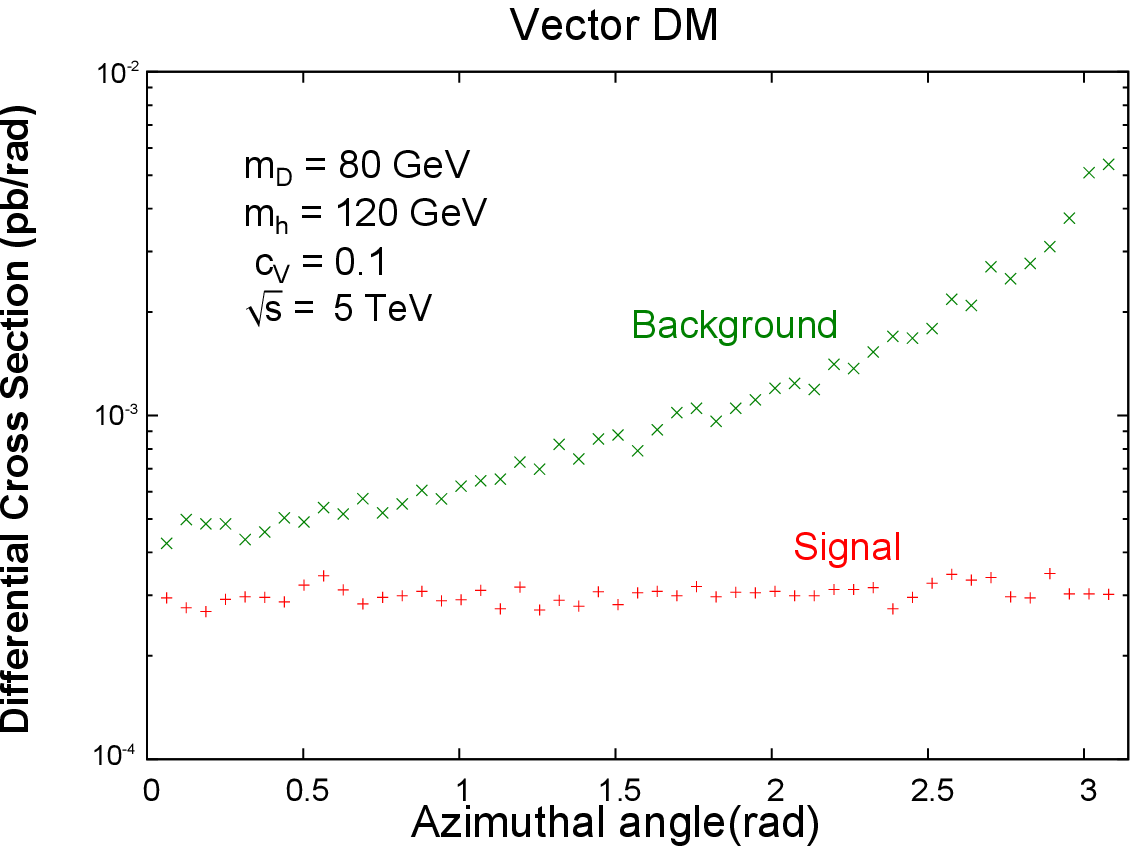}}
\caption{\small Distributions of the missing energy $E_{\rm inv}$ (upper panel), the invariant mass $M_{ee}$ (middle panel), and the azimuthal angle $\phi_{ee}$ between outgoing electron and positron (lower panel) for the signal and the backgrounds events. The center of mass energy is fixed to be $\sqrt{s} = 5$~TeV. The signal (red) is for the vector dark matter with the mass of 80~GeV, while the backgrounds (green) with the final state of $e^+ e^- \nu_e \overline{\nu}_e + e^+ e^- \nu_{\mu} \overline{\nu}_{\mu} + e^+ e^-\nu_{\tau} \overline{\nu}_{\tau}$ are shown. The coupling constant $c_{\rm V}^{}$ is taken to be 0.1.}
\label{fig:cuts}
\end{center}
\end{figure}

First, as seen in the upper panel of Fig.~\ref{fig:cuts}, the signal events tend to be located with lower values of the missing energy $E_{\rm inv}$, while the backgrounds are distributed with larger values. We therefore impose the cut on $E_{\rm inv}$ as
\begin{eqnarray}
 E_{\rm inv}
 <
 0.4\sqrt{s}~{\rm GeV}.
 \label{Eq:Einv}
\end{eqnarray}
By using this cut, the backgrounds from $e^+e^-\nu_{e}\overline{\nu}_e$ can be reduced. 

Second, as seen in the middle panel of Fig.~\ref{fig:cuts}, the invariant mass distributions for outgoing electron and positron are clearly different between the signal and the backgrounds. We therefore impose the kinematical cuts to the invariant mass $M_{ee}$ depending on the spin of the dark matter for $\sqrt{s} = 1$~TeV and 5~TeV as 
\begin{eqnarray}
 M_{ee}^{S1}
 >
 600~{\rm GeV},
 \ \ \ \ 
 M_{ee}^{F1}
 >
 600~{\rm GeV},
 \ \ \ \ 
 M_{ee}^{V1}
 >
 600~{\rm GeV},
 \nonumber \\
 M_{ee}^{S5}
 >
 4200~{\rm GeV},
 \ \ \ \ 
 M_{ee}^{F5}
 >
 3900~{\rm GeV},
 \ \ \ \ 
 M_{ee}^{V5}
 >
 3000~{\rm GeV}.
 \label{Eq:Mee}
\end{eqnarray}

Finally, the distributions of the azimuthal angle $\phi_{ee}$ between outgoing electron and positron are shown for the signal and the background processes in  the lower panel of Fig.~\ref{fig:cuts}. As can be seen in the figure, the signal is insensitive to the azimuthal angle, while most of the background events are located in the region with relatively large values of the azimuthal angle. We therefore impose the cut on $\phi_{ee}$ as 
\begin{eqnarray}
 \phi_{ee}
 <
 2.3~{\rm rad},
 \label{Eq:phi}
\end{eqnarray}
by which a considerable amount of the backgrounds can be eliminated. 

In Table~\ref{table:event}, the event numbers are shown for both the signal and the backgrounds after imposing the above kinematical cuts in Eqs.~(\ref{Eq:Einv})-(\ref{Eq:phi}) in the models with the scalar, fermion, and vector dark matter. Coupling constants are fixed to be $c_{\rm S}^{} = c_{\rm V}^{} = 1$ and $c_{\rm F}^{}/\Lambda = 0.1$~GeV$^{-1}$. We here assume the center of mass energy $\sqrt{s}$ to be 1~TeV and 5~TeV and the integrated luminosity to be 1~ab$^{-1}$. The background from $e^+e^-\nu_e\overline{\nu}_e$ process can be considerably reduced by these kinematical cuts. As a result, the significance to detect the signal, which is defined by 
\begin{eqnarray}
 {\rm Significance} = \frac{N_S}{\sqrt{N_S+N_B}},
 \label{Eq:sig}
\end{eqnarray}
with $N_{S}$ ($N_{B}$) being the event number for signal (backgrounds), can be greater than one even if the Higgs boson cannot decay into a dark matter pair.

\begin{table}[t]
\begin{center}
 \begin{tabular}{cc} 
  \scalebox{0.65}{
  \begin{tabular}{|p{25mm}|@{\vrule width 1.8pt\ }p{15mm}|p{15mm}|p{15mm}|p{15mm}|}
   \hline
   $\sqrt{s}=$1~TeV &basic&$E_{\rm inv}$&$M_{ee}$&$\phi_{ee}$
    \\ \noalign{\hrule height 1.8pt}
    $e^+e^-\phi\phi$&91&77&64&43 
    \\ \noalign{\hrule height 1.8pt}
    $e^+e^-\nu_e\overline{\nu}_e$&115,000&24,600&2,510&791 
    \\ \hline
    $e^+e^-\nu_{\mu}\overline{\nu}_{\mu}$&887&103&61&31 
    \\ \hline
    $e^+e^-\nu_{\tau}\overline{\nu}_{\tau}$&887&103&61&31 
    \\ \hline
   \end{tabular}
   }
&
  \scalebox{0.65}{
  \begin{tabular}{|p{25mm}|@{\vrule width 1.8pt\ }p{15mm}|p{15mm}|p{15mm}|p{15mm}|}
   \hline
    $\sqrt{s}=$5~TeV&basic&$E_{\rm inv}$&$M_{ee}$&$\phi_{ee}$
    \\ \noalign{\hrule height 1.8pt}
    $e^+e^-\phi\phi$&348&342&232&162 
    \\ \noalign{\hrule height 1.8pt}
    $e^+e^-\nu_e\overline{\nu}_e$&148,000&5,150&376&138
    \\ \hline
    $e^+e^-\nu_{\mu}\overline{\nu}_{\mu}$&374&191&53&31 
    \\ \hline
    $e^+e^-\nu_{\tau}\overline{\nu}_{\tau}$&374&191&53&31
    \\ \hline
   \end{tabular}
   }
\\[1.5cm]
  \scalebox{0.65}{
  \begin{tabular}{|p{25mm}|@{\vrule width 1.8pt\ }p{15mm}|p{15mm}|p{15mm}|p{15mm}|}
   \hline
    $\sqrt{s}=$1~TeV&basic&$E_{\rm inv}$&$M_{ee}$&$\phi_{ee}$
    \\ \noalign{\hrule height 1.8pt}
    $e^+e^-\chi\overline{\chi}$&47,800&29,200&23,300&15,300 
    \\ \noalign{\hrule height 1.8pt}
    $e^+e^-\nu_e\overline{\nu}_e$&115,000&24,600&2,510&791 
    \\ \hline
    $e^+e^-\nu_{\mu}\overline{\nu}_{\mu}$&887&103&61&31 
    \\ \hline
    $e^+e^-\nu_{\tau}\overline{\nu}_{\tau}$&887&103&61&31 
    \\ \hline
   \end{tabular}
   }
&
  \scalebox{0.65}{
  \begin{tabular}{|p{25mm}|@{\vrule width 1.8pt\ }p{15mm}|p{15mm}|p{15mm}|p{15mm}|}
   \hline
    $\sqrt{s}=$5~TeV&basic&$E_{\rm inv}$&$M_{ee}$&$\phi_{ee}$
    \\ \noalign{\hrule height 1.8pt}
    $e^+e^-\chi\overline{\chi}$&387,000&361,000&235,000&167,000 
    \\ \noalign{\hrule height 1.8pt}
    $e^+e^-\nu_e\overline{\nu}_e$&148,000&5,150&701&264
    \\ \hline
    $e^+e^-\nu_{\mu}\overline{\nu}_{\mu}$&374&191&80&45 
    \\ \hline
    $e^+e^-\nu_{\tau}\overline{\nu}_{\tau}$&374&191&80&45
    \\ \hline
   \end{tabular}
   }
\\[1.5cm]
  \scalebox{0.65}{
  \begin{tabular}{|p{25mm}|@{\vrule width 1.8pt\ }p{15mm}|p{15mm}|p{15mm}|p{15mm}|}
   \hline
    $\sqrt{s}=$1~TeV&basic&$E_{\rm inv}$&$M_{ee}$&$\phi_{ee}$
    \\ \noalign{\hrule height 1.8pt}
    $e^+e^-V_{\mu}V^{\mu}$&1830&818&649&427 
    \\ \noalign{\hrule height 1.8pt}
    $e^+e^-\nu_e\overline{\nu}_e$&115,000&24,600&2,510&791 
    \\ \hline
    $e^+e^-\nu_{\mu}\overline{\nu}_{\mu}$&887&103&61&31 
    \\ \hline
    $e^+e^-\nu_{\tau}\overline{\nu}_{\tau}$&887&103&61&31 
    \\ \hline
   \end{tabular}
   }
&
  \scalebox{0.65}{
  \begin{tabular}{|p{25mm}|@{\vrule width 1.8pt\ }p{15mm}|p{15mm}|p{15mm}|p{15mm}|}
   \hline
    $\sqrt{s}=$5~TeV&basic&$E_{\rm inv}$&$M_{ee}$&$\phi_{ee}$
    \\ \noalign{\hrule height 1.8pt}
    $e^+e^-V_{\mu}V^{\mu}$&75,500&48,400&43,500&31,500 
    \\ \noalign{\hrule height 1.8pt}
    $e^+e^-\nu_e\overline{\nu}_e$&148,000&5,150&2190&820
    \\ \hline
    $e^+e^-\nu_{\mu}\overline{\nu}_{\mu}$&374&191&147&79 
    \\ \hline
    $e^+e^-\nu_{\tau}\overline{\nu}_{\tau}$&374&191&147&79
    \\ \hline
   \end{tabular}
   }
\\[1.5cm]
\end{tabular}
  \caption{\small Number of events before and after each kinematical cut for the integrated luminosity 1~ab$^{-1}$. We take $c_{\rm S}^{} = c_{\rm V}^{} = 1$ and $c_{\rm F}^{}/\Lambda = 0.1$~GeV$^{-1}$ and $m_D^{}= 80$~GeV.}
  \label{table:event}
 \end{center}
\end{table}

\begin{figure}[t]
 \begin{tabular}{cc}
  \begin{minipage}{0.5\hsize}
   \begin{center}
  \scalebox{0.35}{\includegraphics{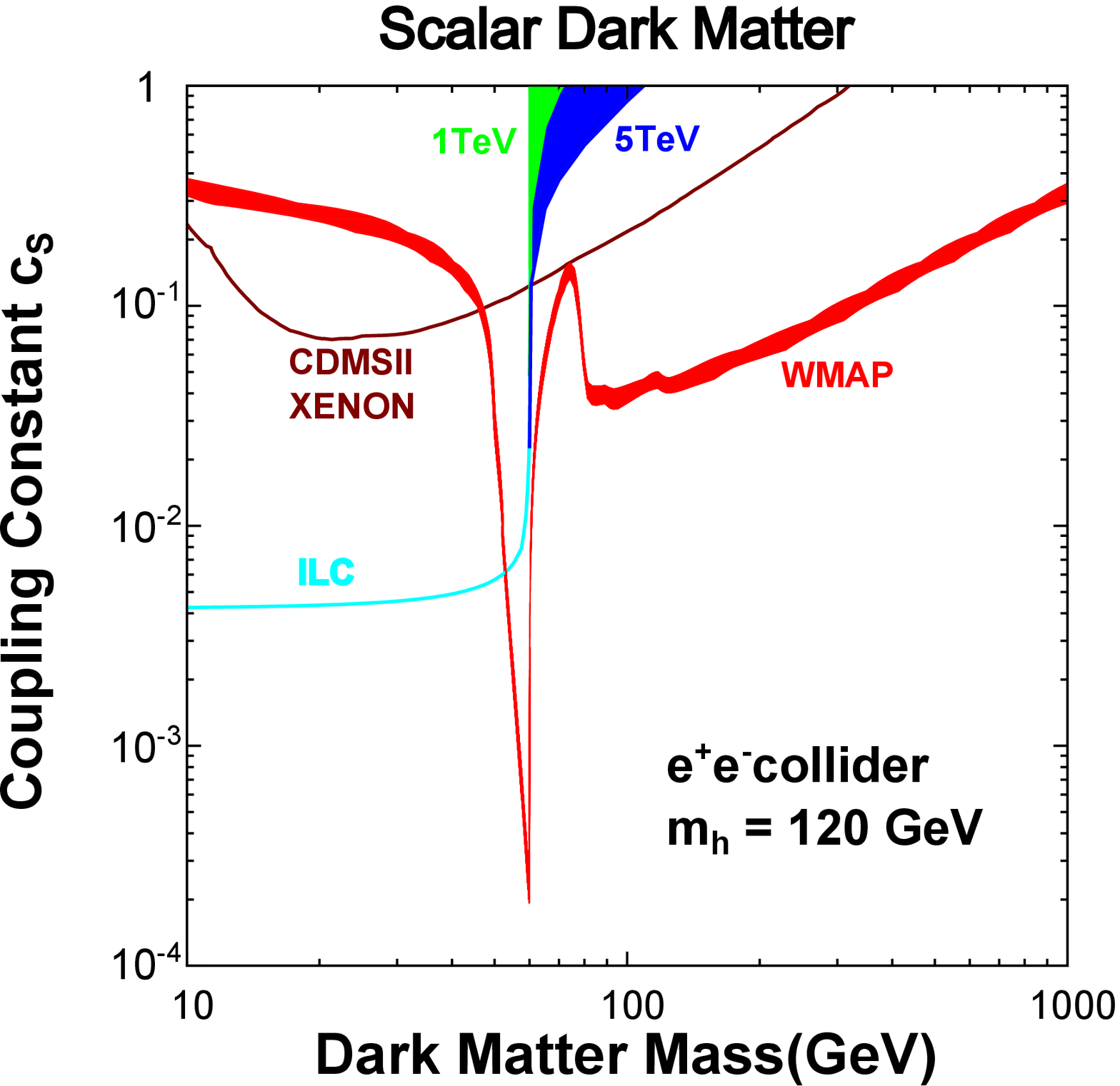}}
  \\
 (a)
%  \label{fig:scalar}
 \end{center}
  \end{minipage}
  \begin{minipage}{0.5\hsize} 
 \begin{center}
 \scalebox{0.35}{\includegraphics{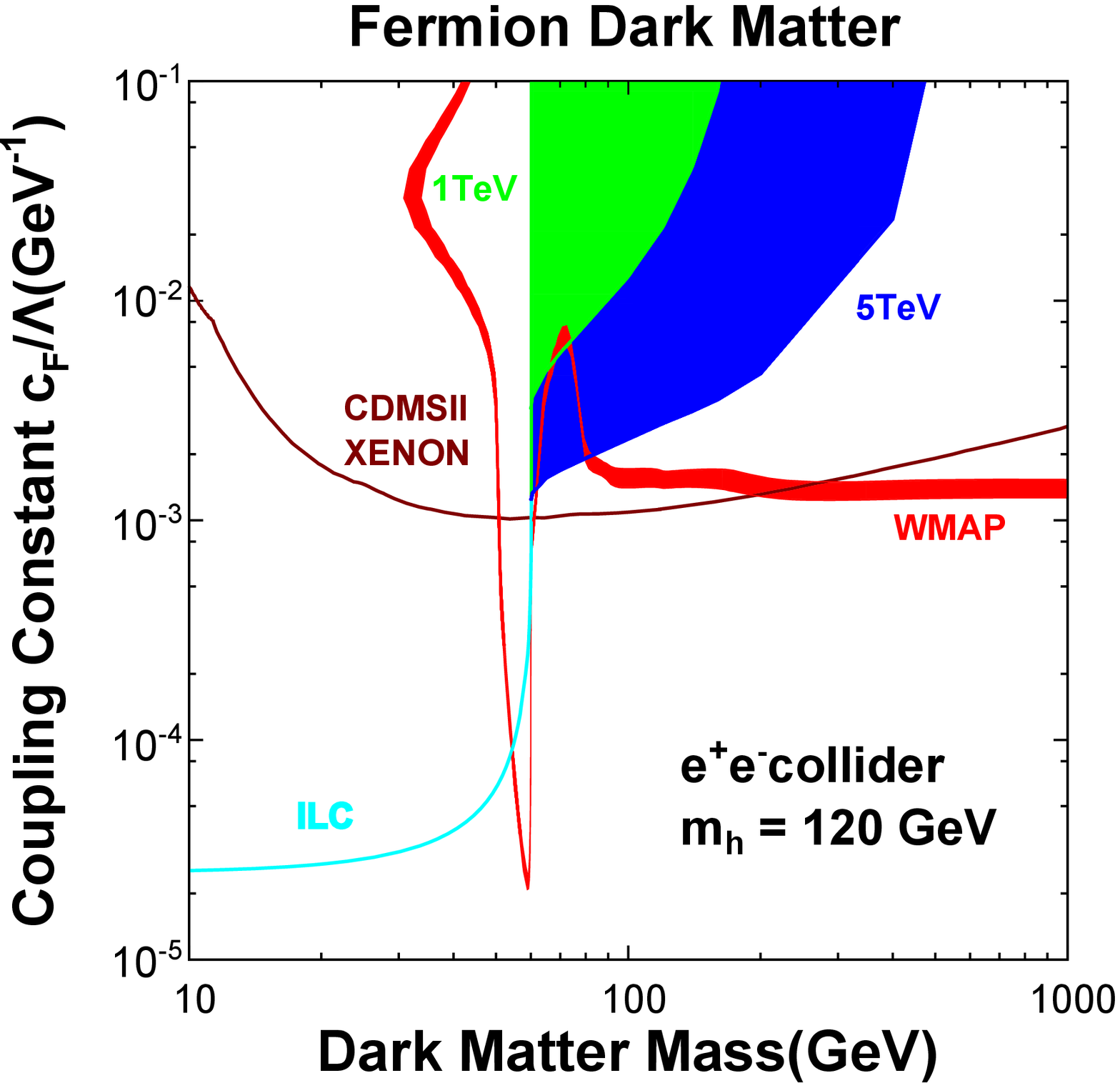}}
 \\
 (b)
%  \label{fig:scalaree}
\end{center}
  \end{minipage}
  \end{tabular}
\\[1.5cm]
%%%%%%%%%%%%%
  \begin{minipage}{0.5\hsize} 
 \begin{center}
  \scalebox{0.35}{\includegraphics{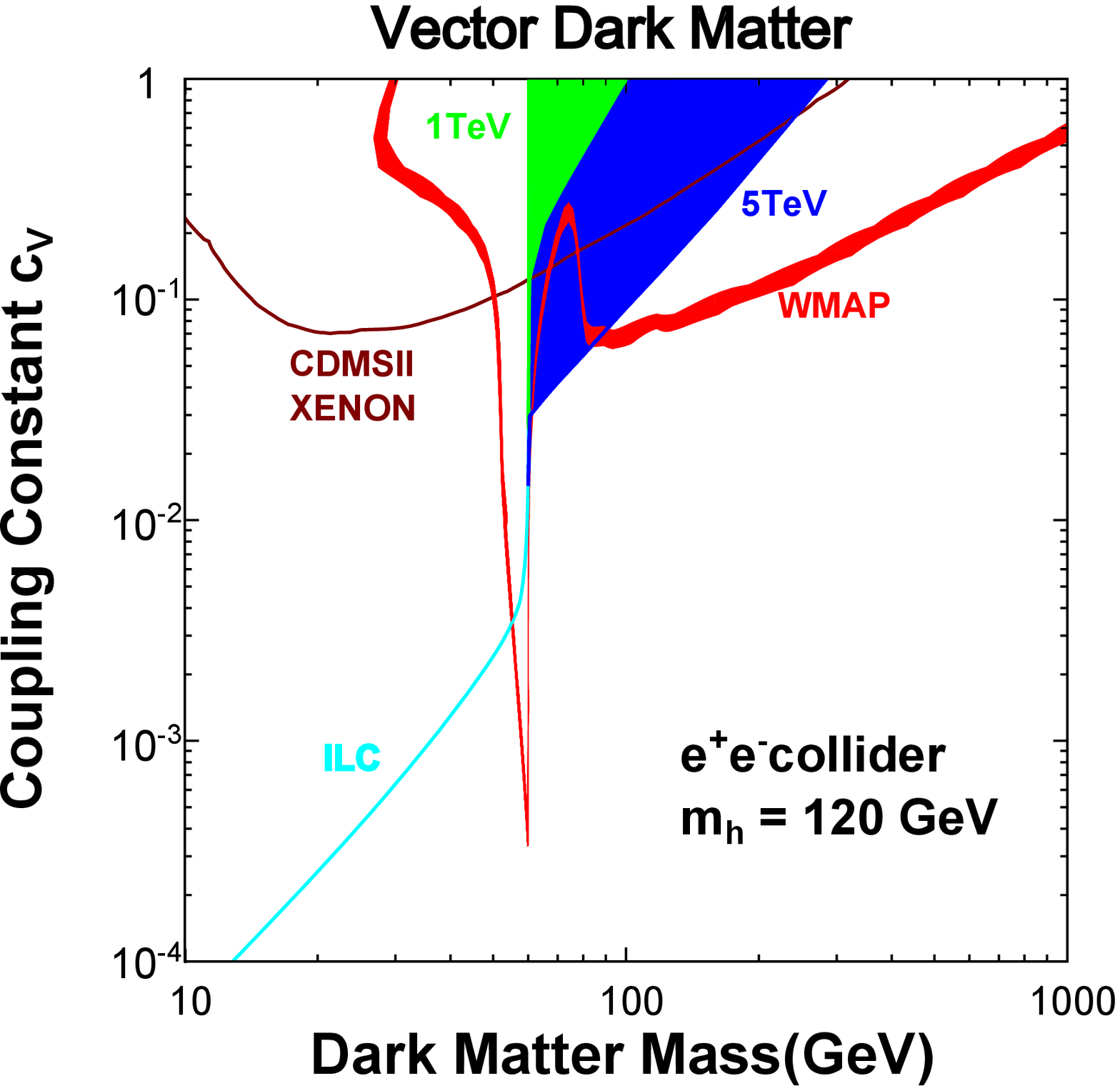}}
 \\
 (c)
%  \label{fig:fermion}
 \end{center}
  \end{minipage}
\caption{\small The areas of $N_S/\sqrt{N_S+N_B} > 3$ at the $e^+e^-$ collider for $\sqrt{s} = 1$~TeV (green) and 5~TeV (blue) with 1 ab$^{-1}$ data are shown with assuming $m_h = 120$~GeV. Parameter regions consistent with the WMAP data (red) and the 90\% C.L. excluded regions by CDMS II and XENON experiments (brown) are also given. For $m_D<m_h/2$, the $3\sigma$ line (cyan) with $\sqrt{s} = 350$~GeV and the integrated luminosity 500~fb$^{-1}$ is shown.}
\label{fig:all1}
\end{figure}

In Figs.~\ref{fig:all1}(a)-\ref{fig:all1}(c), we show the regions where the significance is larger than three in the plane of the coupling constant and the dark matter mass at $\sqrt{s} = 1$~TeV (green area) and 5~TeV (blue area) for $m_D^{} > m_h^{}/2$. The mass of the Higgs boson is set to be 120~GeV and the integrated luminosity is assumed to be 1 ab$^{-1}$. For the region $m_D < m_h/2$, where the Higgs boson can decay into a pair of dark matters, the 3$\sigma$ line at $\sqrt{s} = 350$~GeV with the integrated luminosity 500~fb$^{-1}$ is shown by the cyan curve. In each figure, the allowed region which satisfies the WMAP data (3$\sigma$) is indicated by the red area. We also show the excluded region (90\% C.L.) from direct search results by CDMS~II and XENON~100 by the brown curve\footnote{We are assuming that the scattering cross section is determined only by the diagram in which the Higgs boson is exchanged, Namely, it is not interfered by other diagrams. Furthermore, the detection rate at the direct search has some ambiguities from the hadron matrix element, the dark matter density in the solar system, and the velocity distribution of the dark matter in our galaxy. Those curves should be therefore regarded as a reference.}. 

First, in Fig.~\ref{fig:all1}(a), the results for the scalar dark matter are shown. There is no overlap between the region of $N_S/\sqrt{N_S+N_B} > 3$ and that satisfying the WMAP data even at $\sqrt{s} = 5$~TeV. Second, in Fig.~\ref{fig:all1}(b), the results for the fermion dark matter are shown. For the $e^+e^-$ collision at $\sqrt{s} = 1$~TeV, the area where $N_S/\sqrt{N_S+N_B} > 3$ and the WMAP data are both satisfied is very limited, while the area becomes wider at $\sqrt{s} = 5$~TeV. Finally, in Fig.~\ref{fig:all1}(c), the results for the vector dark matter are shown. At $\sqrt{s} = 1$~TeV, $N_S/\sqrt{N_S+N_B} > 3$ and the WMAP data cannot be compatible, but a wide region of the overlap can be seen at the $\sqrt{s} = 5$~TeV. In particular, for the $m_h/2<m_D<100$~GeV, it can be seen that the vector dark matter with the coupling constant larger than $2$-$4\times 10^{-3}$ can be tested. 

\begin{figure}[t]
 \begin{tabular}{cc}
  \begin{minipage}{0.5\hsize}
   \begin{center}
  \scalebox{0.35}{\includegraphics{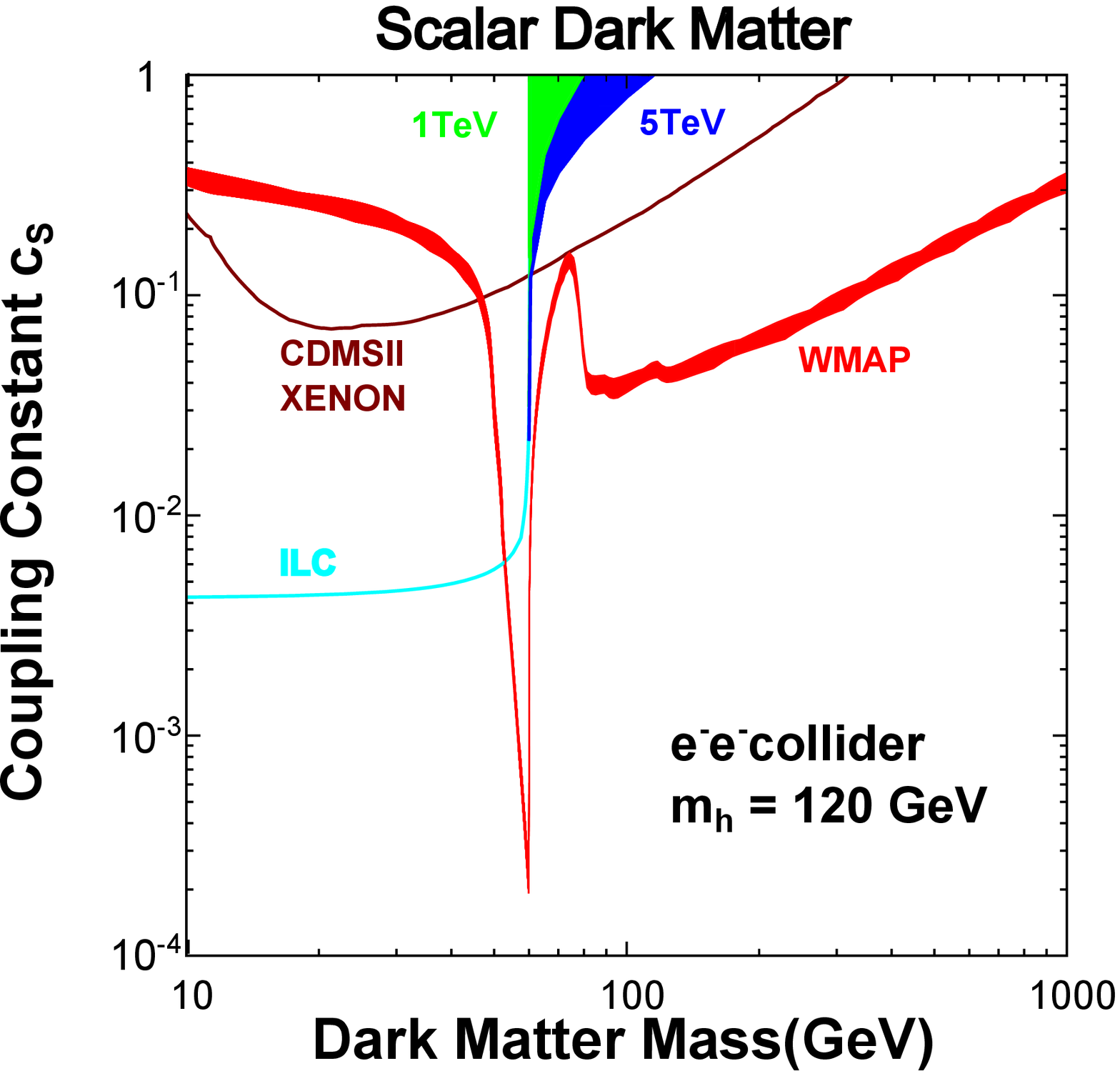}}
  \\
 (a)
%  \label{fig:scalar}
 \end{center}
  \end{minipage}
  \begin{minipage}{0.5\hsize} 
 \begin{center}
 \scalebox{0.35}{\includegraphics{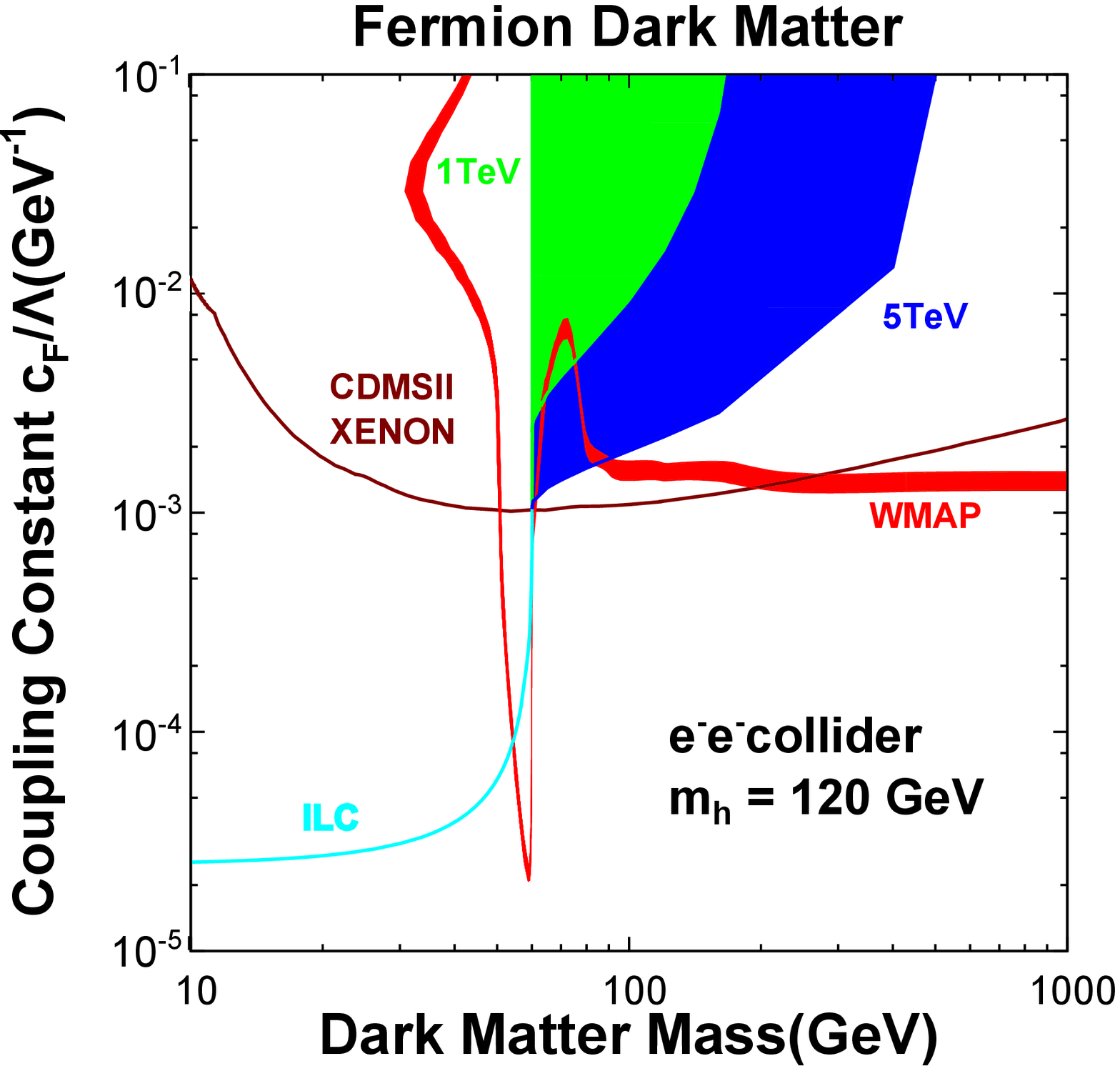}}
 \\
 (b)
%  \label{fig:scalaree}
\end{center}
  \end{minipage}
  \end{tabular}
\\[1.5cm]
%%%%%%%%%%%%%
  \begin{minipage}{0.5\hsize} 
 \begin{center}
  \scalebox{0.35}{\includegraphics{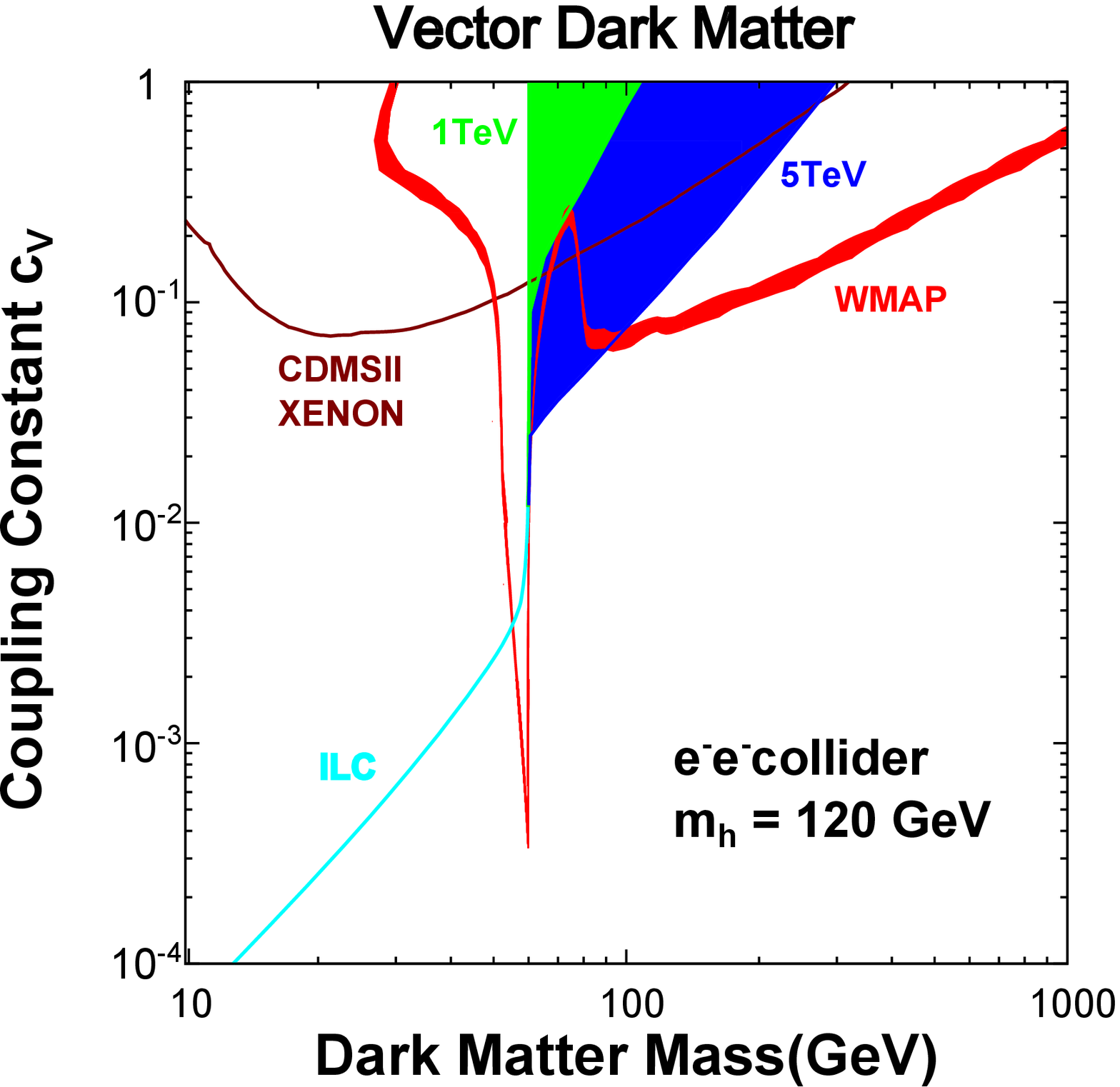}}
  \\
 (c)
%  \label{fig:fermion}
 \end{center}
   \end{minipage}
\caption{\small The areas of $N_S/\sqrt{N_S+N_B} > 3$ at the $e^-e^-$ collider for $\sqrt{s} = 1$~TeV (green) and 5~TeV (blue) with the integrated luminosity of 1 ab$^{-1}$ are shown with assuming $m_h = 120$~GeV. Other lines are the same as those in Fig.~\ref{fig:all1}.}
\label{fig:all2}
\end{figure}

Finally, we discuss the $e^-e^-\to e^-e^-Z^{\ast}Z^{\ast}\to e^-e^-h^{\ast}\to e^-e^- DD$ in electron-electron collisions. The results are very similar to those for the $e^+e^-$ collision. In Fig.~\ref{fig:all2}, we show the results for the $e^-e^-$ collisions, in the model with the scalar, fermion and vector dark matter respectively. For each case, we have imposed the same kinematical cuts as the corresponding processes in the $e^+e^-$ collision. The results in the figure for the $e^-e^-$ collision are qualitatively very similar to those in Fig.~\ref{fig:all1} for the $e^+e^-$ collision. Because of electron-electron collisions, the polarization is more effective than the case of the electron-positron collision.

%%%%%%%%%%%%%%%%%%%%%%%%%%%%%%%%%%
%%%%%%%%%%% Conclusion %%%%%%%%%%%
%%%%%%%%%%%%%%%%%%%%%%%%%%%%%%%%%%
\section{Conclusions}

We have investigated the possibility of detecting dark matter at TeV scale linear colliders in the Higgs portal dark matter scenario with the scalar, fermion or vector dark matter, via $Z$ boson fusion processes at electron-positron and electron-electron collisions. We have found that a multi-TeV collider can be more useful to explore the dark matter in these models than the 1~TeV collider when the invisible decay of the Higgs boson into a pair of dark matters is kinematically forbidden.

\vspace{1.0cm}
%\hspace{0.2cm}
\noindent
{\bf Acknowledgments}
\vspace{0.5cm}

The work was supported in part, by Grant-in-Aid for Science Research, 
Ministry of Education, Culture, Sports, Science and Technology, 
Japan (Nos. 19540277 and 22244031 for SK, and 21740174 and 22244021 for SM),  
and by World Premier International
Research Center Initiative (WPI Initiative), MEXT, Japan.

\end{document}